# Tunable Thermal Conductivity and Mechanical Properties of Metastable Silicon by Phase Engineering

Yubing Du,[1,2,#] Guoshuai Du,[1,2,#] Zhixi Zhu,[3] Jiaohui Yan,[3] Jiayin Li,[1,4] Tiansong Zhang,[1,2] Lina Yang,[2,*] Ke Jin,[1,3] and Yabin Chen[1,2,5,*]

[1]*Advanced Research Institute of Multidisciplinary Sciences (ARIMS), Beijing Institute of Technology, Beijing 100081, P. R. China*

[2]*School of Aerospace Engineering, Beijing Institute of Technology, Beijing 100081, P. R. China*

[3]*School of Materials Science and Engineering, Beijing Institute of Technology, Beijing 100081, P. R. China*

[4]*School of Chemistry and Chemical Engineering, Beijing Institute of Technology, Beijing, 100081 P. R. China*

[5]*BIT Chongqing Institute of Microelectronics and Microsystems, Chongqing, 400030, P. R. China*

[*]Correspondence and requests for materials should be addressed to: chyb0422@bit.edu.cn (Y.C.), and yangln@bit.edu.cn (L.Y.)

[#]These authors contributed equally to this work.

# ABSTRACT

The extensive applications of cubic silicon in flexible transistors and infrared detectors are much hindered by its intrinsic properties. Metastable silicon phases, such as Si-III, IV and XII prepared using extreme pressure method, provide a unique "genetic bank" with diverse structures and exotic characteristics, however, exploration on their inherent physical properties remains immature. Herein, we demonstrate the phase engineering strategy to modulate the thermal conductivity and mechanical properties of metastable silicon. The thermal conductivity obtained via Raman optothermal approach presents the broad tunability across various Si-I, III, XII and IV phases. The hardness and Young's modulus of Si-IV are remarkably greater than those of Si-III/XII mixture, confirmed by nanoindentation technique. Moreover, it was found that the pressure-induced structural defects can substantially degrade the thermal and mechanical properties of silicon. This systematic investigation can offer feasible route to design novel semiconductors and further advance their desirable applications in advanced nanodevices and mechanical transducers.

**INTRODUCTION**

As the key component of integrated circuits and photovoltaics, cubic silicon (Si-I) has aroused extensive attention in semiconductor industry over the past few decades, owing to its superior electronic and optical properties.[1-4] The further applications of Si-I are much impeded by its intrinsic properties, such as the infrared photodetectors and flexible nanodevices.[5] Metastable silicon, synthesized by high-pressure treatment, has emerged as a rational branch to offer the distinguished properties to Si-I.[6,7] It has been found that Si can undergo several phase transitions under hydrostatic or uniaxial pressure, including the metallic Si-II ($\beta$-Sn), Si-XI (Imma), Si-V (hexagonal), Si-VI (Cmca), Si-VII (hcp), and Si-X (fcc) phases.[8-10] Intriguingly, upon decompression, tetragonal Si-II can irreversibly transform to the mixed Si-XII (R8) and Si-III (BC8) phases, rather than the initial Si-I, and the latter can then become hexagonal Si-IV through thermal annealing.[11] Despite considerable efforts, the intrinsic thermal and mechanical properties of these metastable phases remain largely unexplored.[12-17]

Metastable silicon phases provide a versatile platform to modulate their thermal properties and phonon transport behaviors.[18] First−principles calculations combined with Boltzmann transport indicated that the thermal conductivity of rhombohedral Si-XII (16.8 W·m$^{-1}$·K$^{-1}$) is about one order of magnitude lower than that of cubic Si-I (144 W·m$^{-1}$·K$^{-1}$), attributed to its stronger three-phonon scattering.[19] The thermal conductivity of body-centered Si-III, prepared through direct transformation using multi-anvil method, was determined to be 20 W·m$^{-1}$·K$^{-1}$ at room temperature,[16] although the intrinsic physical properties of Si-III still remain contentious. In practice, Si-III/XII mixture normally co-exists when the metallic Si-II is fully decompressed, due to their equivalent potential energy, yielding the variable thermal conductivity from 7.6 to 22.2 W·m$^{-1}$·K$^{-1}$ via time-domain thermoreflectance technique.[20] In addition, the thermal properties of hexagonal Si-IV phase are still unclear, and first-principles calculations showed its thermal conductivity is around 40% lower than cubic Si-I.[21]

Investigations on the mechanical properties of metastable Si phases have stayed on its infant stage, to the best of our knowledge. It has been found that elastic moduli (152 GPa) and hardness (12.0 GPa) of polycrystalline Si-IV phase are similar to those of cubic Si-I, as confirmed by nanoindentation and X-ray diffraction.[22] Notably, the Si-IV phase can display various polytypes depending to their stacking orders, such as 2H (AB stacking) and 4H (ABCB stacking) structures.[14] Moreover, the elastic modulus of high-quality Si crystal can be significantly weakened when numerous structural defects appear, such as grain boundary and quadratic node, which has been theoretically demonstrated with Budiansky's self-consistent method together with a phase mixture model.[23] Apparently, the experimental determination of mechanical properties of metastable Si phases is highly desirable for their potential applications in flexible devices and advanced mechanical sensors.

In this study, both thermal and mechanical properties of metastable Si-III/XII and Si-IV, as well as the decompressed polycrystalline Si-I, have been comprehensively investigated. The measured thermal conductivity by Raman optothermal approach presents wide turnability across different metastable phases. The phase dependence of hardness and elastic modulus of silicon phases has been developed based on continuous stiffness measurements via nanoindentation technique. Moreover, pressure-induced structural defects play a crucial role in the phonon scattering process, resulting in remarkably degraded thermal and mechanical properties. It is anticipated that the discoveries in this work can lay a solid foundation for potential applications of metastable silicon phases.

**RESULTS AND DISCUSSION**

The controlled preparation of metastable Si-III/XII and Si-IV phases can be achieved under high pressure via a diamond anvil cell, as reported in our previous study.[24] Raman optothermal technique has been well developed to measure the thermal conductivity of functional materials, facilitated with its non-invasive and reliable features.[25] According to Fourier's law, the heat flux $\Delta P$, transmitted through an area $A$ of material, is proportional to its thermal conductivity $\kappa$ and negative temperature gradient $dT/dx$, that is, $\Delta P/A = -\kappa\, dT/dx$,[26] where $A = \pi(a/2)^2$ corresponds to the heating area (laser source as heater in our case), and thus temperature gradient is generated. Therefore, thermal conductivity $\kappa$ can be derived as $\kappa = 2/\pi a \cdot \Delta P/\Delta T$. In practice, temperature-dependent Raman shift $\Delta\omega/\Delta T$ as well as laser power-dependent Raman shift $\Delta\omega/\Delta P$ can be obtained, leading to the experimental determination of thermal conductivity via $\kappa = \frac{2\cdot(\Delta\omega/\Delta T)}{\pi a\cdot(\Delta\omega/\Delta P)}$.[27] The spot size $a$ of laser beam can be accurately calibrated based on knife-edge method, as depicted in Figure S1.

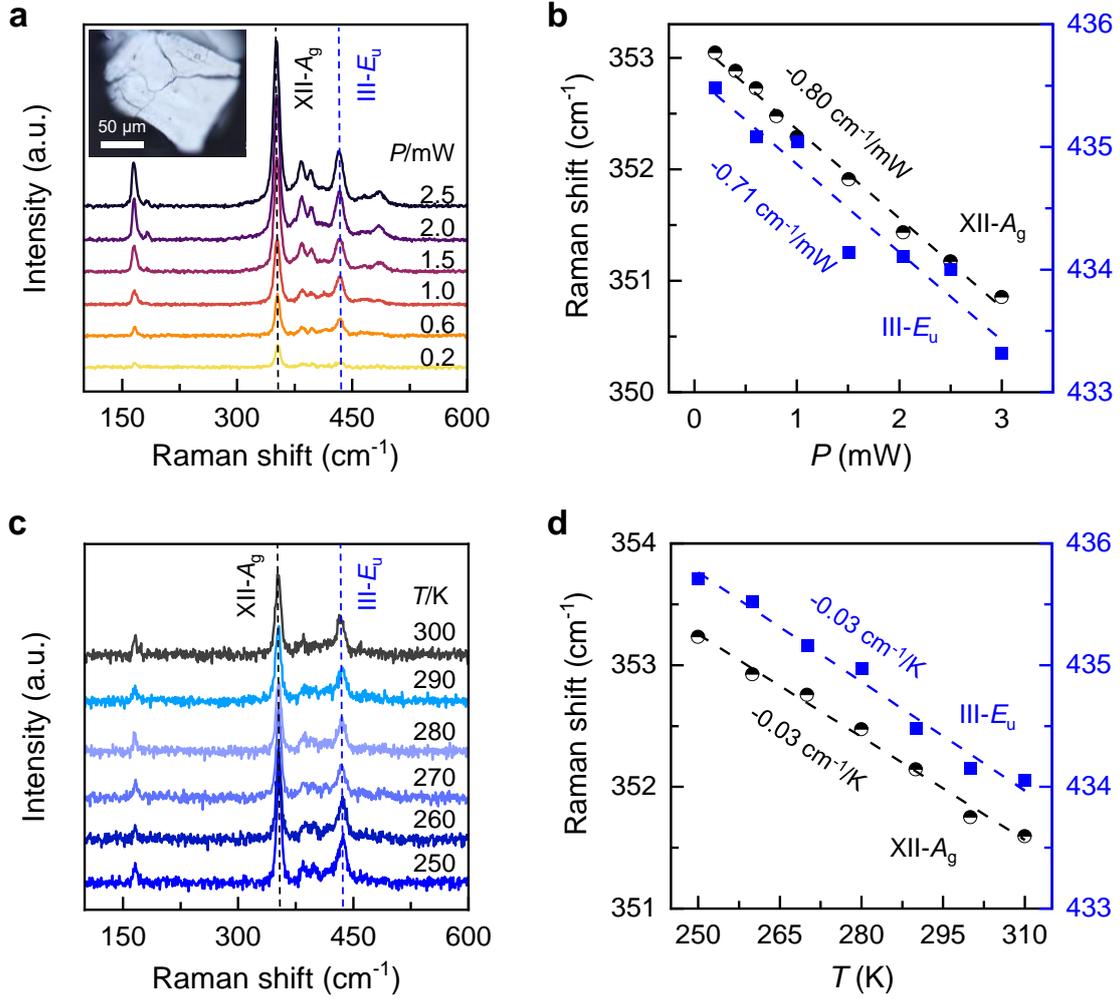

**Figure 1. Thermal conductivity measurement of Si-III/XII mixture based on Raman optothermal technique.** (**a**) The obtained Raman spectra of Si-III/XII with varying laser powers. The inset shows the representative optical image of Si-III/XII sample. The dashed lines indicate the $A_g$ mode of Si-XII (black) and $E_u$ mode of Si-III (blue), observed at 353.0 and 435.5 cm$^{-1}$ under 0.2 mW, respectively. (**b**) Laser power-dependent Raman $A_g$ mode of Si-XII (black) and $E_u$ mode of Si-III phase (blue). The dashed lines mean the linear fitting results. (**c**) Raman spectra of Si-III/ XII measured at different temperatures from 300 K down to 250 K. (**d**) Temperature-dependent Raman $A_g$ mode of Si-XII (black) and $E_u$ mode of Si-III phase (blue). The $I_{\text{Si-III}}/I_{\text{Si-XII}}$ ratio of this measured Si-III/XII sample was around 0.68.

We first investigated the thermal conductivity of the metastable Si-III/XII mixture by using Raman optothermal approach, as shown in Figure 1. For this given metastable sample, the $I_{\text{Si-III}}/I_{\text{Si-XII}}$ was determined to be around 0.68. Following the lattice dynamics and group theory, it was found that Si-III/XII mixture can exhibit two sharp Raman peaks belonging to the $A_g$ mode of Si-XII and $E_u$ mode of Si-III, which appeared at ~353 and ~435 cm$^{-1}$ under low power excitation in our case, respectively. In Figure 1a, the intensity of each Raman peak becomes remarkably stronger as the laser power rises. More importantly, all Raman peaks show a significant redshift with laser power, due to laser-induced heating effect. The linear slope d$\omega$/d$P$ of Raman shifts as a function of laser power was fitted as -0.80 and -0.71 cm$^{-1}$/mW for Si-XII $A_g$ and Si-III $E_u$ modes in Figure 1b, respectively. Within the laser power range explored in this experiment, the maximum Raman shifts approached to ~2.2 and ~2.1 cm$^{-1}$ for Si-XII and Si-III phase, respectively, indicating a local temperature rise of about 80 K.[28] In principle, the phonon energy of silicon lattice becomes weakened under high temperature, owing to the prominent thermal expansion and anharmonic characteristic of interatomic potential. This phenomenon occurred in Figure 1c, where the sample temperature varied from 310 to 250 K in a liquid-nitrogen chamber. Moreover, Raman shifts of both Si-XII $A_g$ and Si-III $E_u$ modes vary linearly within this temperature range. Therefore, the linear fitting process is reasonable, leading to a similar slope d$\omega$/d$T$ of –0.03 cm$^{-1}$/K for Si-XII $A_g$ and Si-III $E_u$ modes, as displayed in Figure 1d. As discussed above, the phenomenological thermal conductivities $\kappa_{\text{Si-XII}}$ and $\kappa_{\text{Si-III}}$ were calculated to be 33.2 and 37.4 W·m$^{-1}$·K$^{-1}$, respectively.

Furthermore, the nominal thermal conductivities of multiple Si-III/XII mixtures with various $I_{\text{Si-III}}/I_{\text{Si-XII}}$ ratios were systematically measured, in order to elucidate its dependence on phase concentration. It is worth noting that the total thermal conductivity of a mixture or alloy is reasonably lower than that of each pure constituent due to the broken phonon coherence and its reduced mean free path.[29,30] As established in our earlier study,[24] the dynamic decompression process plays a great role on the modulation of Si-III to Si-XII ratio. As shown in Figure S2, the phenomenological thermal conductivities of both Si-III and Si-XII phases monotonically decrease as the proportion of Si-III phase grows up, potentially suggesting the greater thermal conductivity of pure Si-XII than that of Si-III. First−principles calculations predict that anisotropic thermal conductivity of pure Si-XII is 16.8 and 8.9 W·m$^{-1}$·K$^{-1}$ along $x$ and $z$ axes, respectively,[19] while the measured thermal conductivity of Si-III phase is 20 W·m$^{-1}$·K$^{-1}$ at 300 K,[16] which are close to those of Si-III/XII mixtures with $I_{\text{Si-III}}/I_{\text{Si-XII}}$ over ~0.8 in our case.

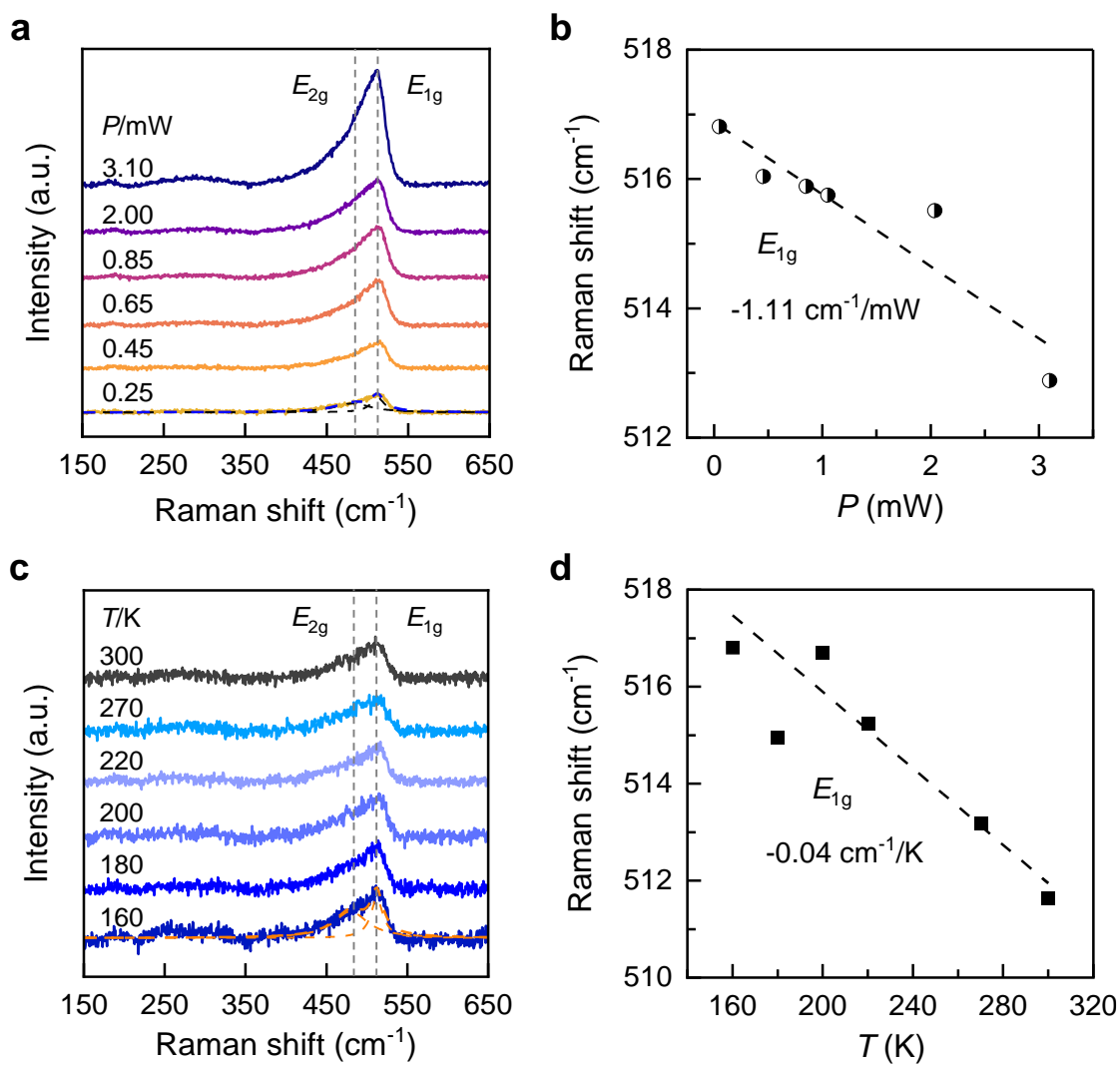

**Figure 2. Thermal conductivity measurement of Si-IV phase based on Raman optothermal technique.** (**a**) Raman spectra of Si-IV at various power levels. (**b**) The relationship between Raman $E_{1g}$ mode of Si-IV and laser power. (**c**) Raman spectra of Si-IV acquired at different temperatures. (**d**) The relationship between Raman $E_{1g}$ mode of Si-IV and temperature. The dashed lines refer to the linear fitting results.

Then, we turned to investigate the thermal conductivity of Si-IV phase using Raman optothermal approach, as shown in Figure 2. The pure Si-IV phase was efficiently obtained by thermal annealing of Si-III/XII, and further confirmed by the synchrotron X-ray diffraction measurement (Figure S3).[24] Figure 2a exhibits the acquired Raman spectra of Si-IV under variable laser powers up to 3.10 mW. For each curve, there is only one asymmetric and broad Raman band appeared, which can be well fitted with two Lorentzian peaks as ~482 and 517 cm$^{-1}$ (0.25 mW), contributing to the $E_{2g}$ and $E_{1g}$ phonon modes of Si-IV phase.[31] The latter $E_{1g}$ mode was selected to analyze the thermal conductivity of Si-IV phase, due to its relatively stronger intensity. The correlation between $E_{1g}$ energy and laser power was found to approximate a linear function with d$\omega$/d$P$ ~ -1.11 cm$^{-1}$/mW as displayed in Figure 2b. Similarly, Figure 2c-d presents the linear dependence of Raman shift on temperature, and the linear slope d$\omega$/d$T$ was extracted as -0.04 cm$^{-1}$/K. Therefore, the thermal conductivity of Si-IV phase was calculated to be 31.9 W·m$^{-1}$·K$^{-1}$, superior to those of Si-III and Si-XII phases, especially for $I_{Si-III}/I_{Si-XII}$ > ~ 0.75 in Figure S2. Notably, our experimental result is obviously lower than the theoretical value of 74 W·m$^{-1}$·K$^{-1}$ at 300 K.[21] This discrepancy arises due to the polycrystalline nature of our Si-IV sample, where the structural defects, such as grain boundary and dislocations induced under high pressure, can dramatically scatter the phonon transport and hence degrade its thermal conductance.[32]

Thermal conductivity tests were systematically carried out on Si-III, Si-XII and Si-IV phases together with the pristine and high-pressure treated Si-I samples, tentatively to uncover the underlying mechanism. The obtained thermal conductivities of metastable phases are summarized in Figure 3, combined with literature data.[16,19,21,33-35] It is obvious that high-pressure treatment can dramatically reduce the thermal conductivity of pristine Si-I, from the initial 139 to 9.6 W·m$^{-1}$·K$^{-1}$ after 8 GPa as an example (the raw data shown in Figures S4-S6). Moreover, these comparative analyses also confirm that the relatively low thermal conductivity to Si-III/XII and Si-IV phases transformed from Si-I under high pressure. In principle, the thermal conductivity of metastable Si phases is governed by phonon transport behavior, and their electronic contribution can be negligible, due to their semiconducting characteristic. Numerous structural defects emerge in silicon lattice under extreme pressure and can be even retained to ambient conditions, which stimulate numerous scattering centers with lattice phonons. Based on the modified Callaway model, we know that phonon lifetime $\tau$ is rapidly shortened by their inelastic interactions with point ($\tau^{-1}$~$\omega^4$), dislocation ($\tau^{-1}$~$\omega^3$) and boundary ($\tau^{-1}$~$\omega^0$) defects, where $\omega$ is the phonon angular frequency.[36] Importantly, thermal annealing can efficiently reconstruct the atomic arrangement, thereby eliminating lattice defects, resulting in the gradually enhanced thermal conductivity. This phenomenon was evidently demonstrated in our annealed Si-I at 1273 K for 10 hours, with its thermal conductivity raised to 13.4 W·m$^{-1}$·K$^{-1}$.

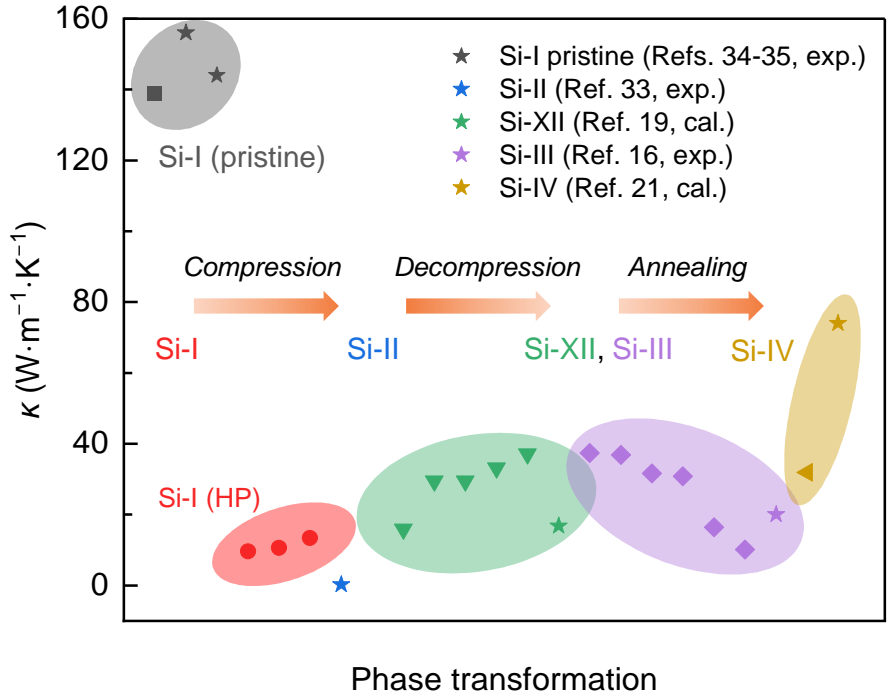

**Figure 3. Thermal conductivity of Si with various phases, including Si-I, Si-XII, Si-III and Si-IV.** The inset illustrates the transition pathways of silicon. The data points marked with stars represent the literature results.

Next, we turn to examine mechanical hardness $H$ and Young's modulus $E_Y$ of metastable Si phases using nanoindentation method,[37] including Si-III/XII mixture and Si-IV compared with pristine Si-I and the decompressed Si-I from 8 GPa. The representative load-displacement curves of each phase in Figure S7 present good repeatability, indicating the reliable mechanical results. The extracted hardness and Young's modulus results are shown in Figure 4. The mean $H$ and $E_Y$ of pristine Si-I phase lie in 12.1 ± 0.5 and 183.5 ± 5.2 GPa, respectively, well consistent with literature data.[38] Once it was subjected to high pressure, the mechanical properties of the Si-I phase were remarkably degraded, like $E_Y$~138.3 ± 7.1 after 8 GPa. This phenomenon can be well explained by the pressure-induced lattice defects.[23] In comparison, the hardness and Young's modulus of Si-III/XII and Si-IV are approximately 35% and 15% lower than those of Si-I phase, respectively. In detail, all Si allotropes are covalently connected with $sp^3$ hybridized bonds. According to Rietveld refinements of XRD results, both hexagonal Si-IV (2.33 and 2.37 Å) and body-centered Si-III (2.34 and 2.39 Å) lattices have two distinct Si-Si bond lengths, whereas rhombohedral Si-XII phase shares four longer bond lengths as 2.38, 2.40, 2.34, and 2.38 Å. Therefore, the comparable modulus of Si-IV and Si-I originate from their similar atomic bond characteristics, both of which are stronger than that of Si-III/XII mixture.

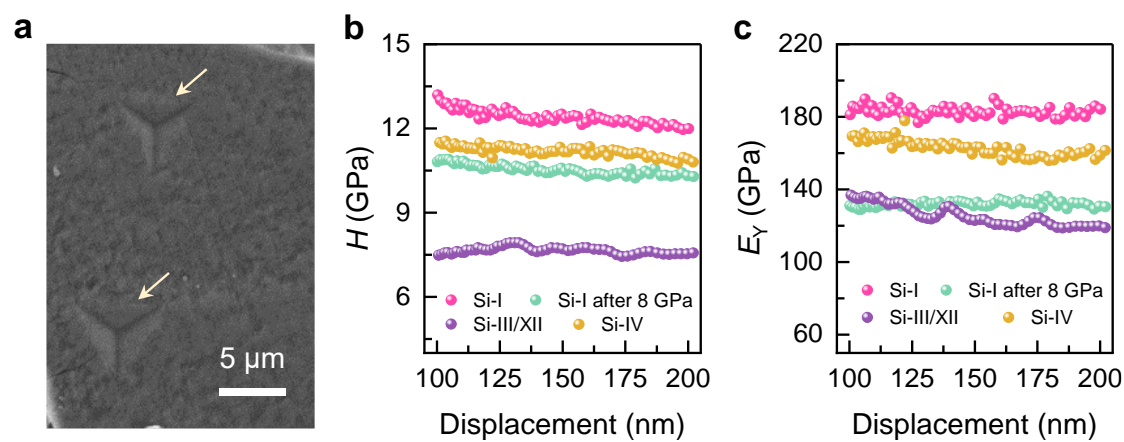

**Figure 4. Hardness and Young's modulus measurements of various silicon phases based on nanoindentation technique.** (**a**) The representative scanning electron microscope image of Si-III/XII after nanoindentation test. The indented areas are obvious as marked with two arrows. (**b**) Relationship between hardness $H$ and displacement for single-crystalline Si-I, polycrystalline Si-I, Si-III/XII, and Si-IV phases. (**c**) Relationship between Young's modulus $E_Y$ and displacement for single-crystalline Si-I, polycrystalline Si-I, Si-III/XII, and Si-IV phases.

We further summarized the hardness and Young's modulus of various Si phases in Figure 5. Clearly, the mechanical properties of Si present a noteworthy tunability, depending on the specific phase structure. For instance, the hardness can be varied from 7.4 ± 0.7 (Si-III/XII) to 12.1 ± 0.5 GPa (Si-I), enabling the fabrication of all Si-based mechanical sensors. Our obtained hardness and elastic moduli results of Si-IV phase are quantitatively comparable with those of cubic Si-I, consistent with the literature data,[22] where the bulk modulus ~91.8 GPa and Poisson's ratio ~0.22 were derived as well. All metastable Si phases present the brittle characteristics, according to their robust covalent bonds. Nanoindentation process induces the irreversible plastic deformation on the silicon surface, as illustrated in Figure 4a. The structural defects caused by extreme pressure can remarkably weaken the mechanical strength of silicon due to pronounced surface and size effects, as evidenced by our Si-I phase, which cannot be intuitively predicated by standard continuum mechanics.[39,40]

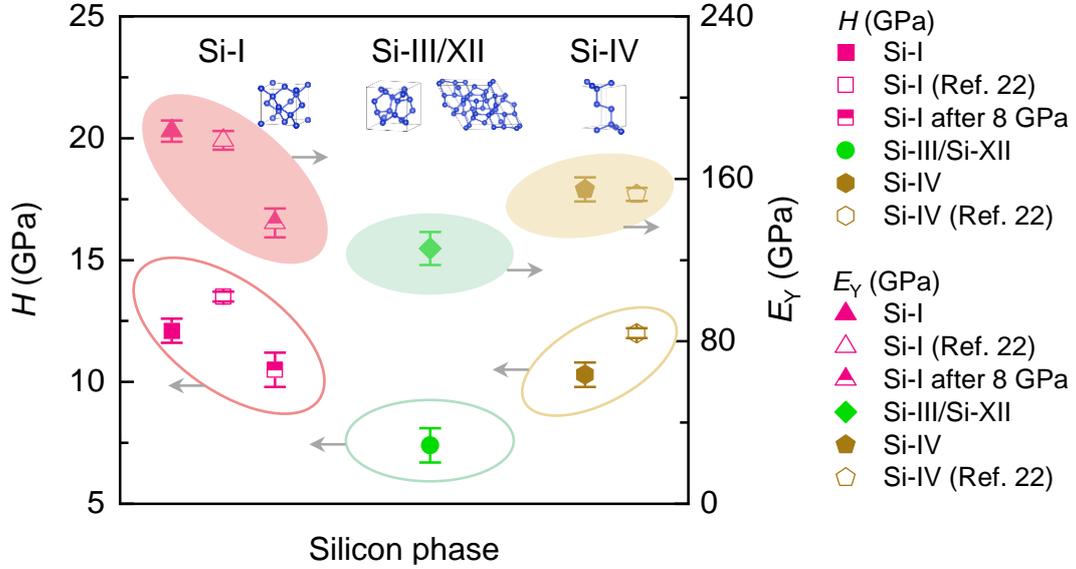

**Figure 5. Phase structure-dependent hardness $H$ and Young's modulus $E_Y$ of silicon.** The inset shows the lattice structure of each silicon phase.

## CONCLUSION

In summary, phase structure-regulated thermal and mechanical properties of metastable silicon have been systematically investigated. It was found that thermal conductivities of Si-III and XII are greatly sensitive to its phase concentration, and thermal conductivity of polycrystalline Si-IV (31.9 W·m$^{-1}$·K$^{-1}$) is much lower than that of Si-I, partially due to the prominent phonon-defect scattering effect. Nanoindentation characterizations revealed that the hardness and Young's modulus of Si-III/XII ($H$~7.4 ± 0.7 GPa and $E_Y$~125.7 ± 8.1 GPa) and Si-IV ($H$~10.3 ± 0.5 GPa and $E_Y$~154.8 ± 5.9 GPa) phases are significantly weaker than those of cubic Si-I phase. These findings are expected to open new avenues for exploring the distinguished properties of novel Si allotropes and further expand their practical applications beyond the typical cubic Si-I counterpart.

## EXPERIMENTAL METHODS

**Raman optothermal approach:** Thermal conductivity of the metastable Si samples was measured based on the optothermal Raman technique, where the laser excitation plays an essential role of heat source. In brief, both temperature dependent-Raman shift d$\omega$/d$T$ and lase power-dependent Raman shift d$\omega$/d$P$ for one specific phonon mode of silicon are fundamental parameters to derive its thermal conductivity. In our case, all Raman spectra of silicon specimens were collected using a homemade Raman spectroscopy equipped with an iHR550 spectrometer (Horiba, 1800 gr/mm grating) and 50× objective lens (NA~0.75). The wavelength of laser source was 532 nm, and its power density was accurately measured using a power meter (Thorlabs, S121C). Each Raman spectrum was acquired after the silicon sample reached thermal equilibrium with its surroundings. The laser power-dependent Raman measurements were performed at room temperature, and the laser power was tuned from 0.05 to 3.10 mW. The temperature-dependent Raman experiments were carried out in a homemade cryogenic setup, providing a wide base temperature range from 140 to 300 K in liquid nitrogen chamber. Meanwhile, the laser power was optimized to be below 0.4 mW, in order to minimize laser-induced heating effect. The typical integration time for each Raman curve was 50 s for Si-III/XII and 80 s for Si-IV.

**Nanoindentation technique:** Nanoindentation measurements of the metastable Si phases were performed using a Nanoindenter apparatus (KLA G200) equipped with a Berkovich probe (correction factor $\beta$~1.034), based on the continuous stiffness measurement technique. The maximum indentation depth was maintained to be far below 1/20 of the sample thickness, and the loading/unloading strain rate of 0.05 s$^{-1}$ was applied throughout the experiments. The allowable thermal drift rate was 0.05 nm/s, and the actual thermal drift was measured and corrected at 90% unloading accordingly.

**Synchrotron XRD measurement:** Crystal structures of the synthesized Si samples were characterized using the advanced synchrotron XRD at the 15U1 beamline of the Shanghai Synchrotron Radiation Facility (SSRF). The wavelength of X-ray beam was 0.6199 Å and it spot size was focused within ~3 μm. Two-dimensional diffraction images were captured with a MAR165 CCD detector and subsequently integrated into one-dimensional XRD profiles using the Dioptas software.[41] The XRD data were further analyzed and refined using the Rietveld method implemented in GSAS-II.[42]

## ASSOCIATED CONTENT

**Supporting Information**

The Supporting Information is available free of charge online.

The spot size measurement of laser beam based on knife-edge approach; Synchrotron X-ray diffraction characterizations of Si-IV and Si-III/ XII; Detailed Raman spectra of Si-I, treated by different pressures and temperatures, under various laser powers and temperatures; The representative load-displacement curves of Si-III/XII and Si-IV phases for mechanical measurements.

**Author Contributions**

Y.C. and Y.D. conceived this research project and designed the experiments. Y.D., G.D. and T.Z. prepared the silicon samples and carried out thermal conductivity and mechanical measurements. K.J., Z.Z. and J.Y. carried out the nanoindentation measurements. Y.D., J.L. and G.D. attributed to the Raman measurements. Y.D. and L.Y. performed all thermal analysis. Y.C. and Y.D. wrote the manuscript with the essential input of other authors. All authors have given approval of the final manuscript.

**Notes**

The authors declare no competing financial interest.

**Acknowledgments**

This work was financially supported by the National Natural Science Foundation of China (grant numbers 52472040, 52072032, and 12090031), and the 173 JCJQ program (grant No. 2021-JCJQ-JJ-0159). We gratefully thank Dr. H.L. Dong for his support on the synchrotron XRD measurements.

**Data Availability**

All data related to this study are available from the corresponding authors upon reasonable request.